\pdfoutput=1
\documentclass[10pt, twocolumn]{article}

\newcommand{\blist}{\begin{easylist}[itemize]}
\newcommand{\elist}{\end{easylist}}
\newcommand{\alpaka}{Alpaka\xspace}
\newcommand{\cuda}{{CUDA}\xspace}
\newcommand{\simd}{{SIMD}\xspace}
\newcommand{\openmp}{{OpenMP}\xspace}
\newcommand{\opencl}{{OpenCL}\xspace}
\newcommand{\cublas}{{cuBlas}\xspace}

\newcommand{\nvidia}{{NVIDIA}\xspace}
\newcommand{\intel}{{Intel}\xspace}
\newcommand{\amd}{{AMD}\xspace}
\newcommand{\github}{{GitHub}\xspace}
\newcommand{\thrust}{{Thrust}\xspace}
\newcommand{\cpp}[1]{\lstinline[identifierstyle=\color{black}\bfseries]{#1}}
\newcommand{\cmark}{\ding{51}}
\newcommand{\xmark}{\ding{55}}
\newcommand{\omark}{\ding{109}}
\newcommand{\issupp}{\cellcolor{green!25}\cmark}
\newcommand{\nosupp}{\cellcolor{red!25}\xmark}
\newcommand{\lisupp}{\cellcolor{yellow!25}\omark}
\newcommand{\naive}{na\"ive\xspace}
\newcommand{\ignore}[1]{}
\newcommand\Mark[1]{\textsuperscript#1}

\usepackage[table]{xcolor}
\usepackage[
colorinlistoftodos,
textsize=scriptsize,
]{todonotes}
\usepackage[sharp]{easylist}
\usepackage{listings}
\usepackage[
  bookmarks=false,
  pdfpagelabels=false,
  hyperfootnotes=false,
  hyperindex=false,
  pageanchor=false,
  hidelinks,
]{hyperref}
\usepackage{amsmath}
\usepackage{parcolumns}
\usepackage{pifont}
\usepackage{xspace}

\definecolor{mygreen}{rgb}{0,0.6,0}
\definecolor{mygray}{rgb}{0.5,0.5,0.5}
\definecolor{mymauve}{rgb}{0.58,0,0.82}
\definecolor{myblue}{HTML}{E5F2FF}

\lstdefinestyle{customptx}{
	numberstyle=\tiny\color{black},
	breaklines=true,
	basicstyle=\footnotesize\ttfamily,
	keywordstyle=\bfseries\color{blue}\ttfamily,
	stringstyle=\color{brown}\ttfamily,
	commentstyle=\itshape\color{mygreen}\ttfamily,
	belowcaptionskip=1\baselineskip,
	tabsize=4,
	xleftmargin=\parindent,
	language=TeX,
	showstringspaces=false,
	identifierstyle=\bfseries\color{myblue},
}

\begin{document}

\title{{\alpaka} -- An Abstraction Library for Parallel Kernel Acceleration\protect\footnote{This project has received funding from the European Union’s Horizon 2020 research and innovation programme under grant agreement No 654220}}

\author{Erik Zenker\Mark{{1,2}}, Benjamin Worpitz\Mark{{1,2}}, Ren\'{e} Widera\Mark{1}, Axel Huebl\Mark{{1,2}},
  \\
  Guido Juckeland\Mark{{1,2}}, Andreas Kn\"{u}pfer\Mark{2}, Wolfgang E. Nagel\Mark{2}, Michael Bussmann\Mark{1}
  \\
  \\
  \Mark{1}Helmholtz-Zentrum Dresden - Rossendorf, Dresden, Germany\\
  \{e.zenker, r.widera, a.huebl, g.juckeland@hzdr.de, m.bussmann\}@hzdr.de
  \\
  \Mark{2}Technische Universit\"at Dresden, Dresden, Germany\\%
  \{andreas.knuepfer, wolfgang.nagel\}@tu-dresden.de, benjamin.worpitz@outlook.com%
}

\maketitle

\begin{abstract}
Porting applications to new hardware or programming models is a tedious and error prone process.
Every help that eases these burdens is saving developer time that can then be invested into the advancement of the application itself instead of preserving the status-quo on a new platform.

The \alpaka library defines and implements an abstract hierarchical redundant parallelism model.
The model exploits parallelism and memory hierarchies on a node at all levels available in current hardware.
By doing so, it allows to achieve platform \emph{and} performance portability across various types of accelerators by ignoring specific unsupported
levels and utilizing only the ones supported on a specific
accelerator.
All hardware types (multi- and many-core CPUs, GPUs and other accelerators) are supported for and can be programmed in the same way.
The \alpaka C++ template interface allows for straightforward extension of the library to support other accelerators and specialization of its internals for optimization.

Running \alpaka applications on a new (and supported) platform requires the change of only one source code line instead of a lot of \#ifdefs.

\end{abstract}


\section{Introduction}

\subsection{Motivation}
Performance gain by employing parallelism in software nowadays faces a variety of obstacles.
Parallel performance currently relies on the efficient use of many-core architectures that are commonly found in a heterogeneous environment of multi-core CPU and many-core accelerator hardware.

Heterogeneous systems often expose a memory hierarchy that has to be used efficiently as high computational performance usually requires high memory throughput, demanding the development of efficient caching strategies by application developers.

The same developers face a variety of parallel computing models either specific to a certain hardware or with limited control over optimization.
Many models aim for providing easy to learn interfaces that hide the complexities of memory management and parallel execution while promising performance portability, but ultimately fall short of at least one of their aims.

Due to the Herculean effort associated with maintaining a multi-source application even large development teams thus usually have to choose a strategy of trading performance for portability or vice versa by choosing one single programming model.

Alpaka was designed to prevent this trade off by providing a single source abstract interface that exposes all levels of parallelism existent on today's heterogeneous  systems.
Alpaka heavily relies on existing parallelism models, but encapsulates them via a redundant mapping of its abstract parallelization hierarchy~\cite{rocki2014future} to a specific hardware, allowing for mixing various models in a single source C++ code at runtime. Thus, hardware-specific optimizations are possible without the necessity for code replication. 

Alpaka therefore is open to future performance optimizations while providing portable code.
This is only possible as Alpaka relies on the developers ability to write parallel code by explicitly exposing all information useful for defining parallel execution in a heterogeneous environment rather than hiding the complexities of parallel programming.

Moreover, Alpaka limits itself to a simple, pointer based memory model that requires explicit deep copies between memory levels.
This puts the developer in the position to develop portable parallel caching strategies without imposing restrictions on the memory layout.
Thus, developers achieve performance portability by skillful code design for which Alpaka provides a single source, explicit redundant parallelism model without any intrinsic optimization hidden from the user. 

In the following, we define some categories in order to compare Alpaka to existing models for parallel programming.

\textbf{Openness} By \emph{Openness} we refer to models licensed as open source or defined by an open standard.

\textbf{Single source} A model that provides for \emph{single source} code allows for the application code to be written in a single programming language. It furthermore does not require extensive multiple compilation branches with varying implementations of an algorithm specific to a certain hardware. \emph{Single source} models may provide annotations (e.g. compiler directives) or add defined words to the language that describe parallel execution.

\textbf{Sustainability} We define a \emph{sustainable} parallel programming model as a model where the porting of an algorithm to another hardware requires minimum changes to the algorithmic description itself. \emph{Sustainable} models furthermore should be adaptable to future hardware and be available for at least two varieties of current hardware architectures.

\textbf{Heterogeneity} Parallel programming models map to \emph{heterogeneous} systems if they allow for developing a \emph{single source} code in such a way that execution on various hardware architectures requires minimum specific changes (e.g. offloading, memory scope), execution of a single algorithmic implementation on various architectures can happen in the same program and at the same time during run time.

\textbf{Maintainability} We define a parallel programming model to serve code \emph{maintainability} if it provides a \emph{single source} code that is \emph{sustainable} and allows for execution on \emph{heterogeneous} hardware by changing or extending the programming model rather than the application source code.

\textbf{Testability} A model provides \emph{testability} if an algorithmic implementation can be tested on a specific hardware and give, in a lose sense, the same results when migrating to another hardware. \emph{Testability} requires \emph{sustainability}, \emph{heterogeneity} and \emph{maintainability} but furthermore demands a separation of the algorithmic description from hardware specific features.

\textbf{Optimizability} We define an \emph{optimizable} model by the fact that it provides the user with complete control over the parallelization of the algorithm as well as the memory hierarchy in a \emph{heterogeneous} system. Furthermore, fine-tuning algorithmic performance to a specific hardware should not force developers to write multiple implementations of the same algorithm, but rather be provided for by the model.

\textbf{Data structure agnostic} A \emph{data structure agnostic} model does not restrict the memory layout, it instead provides full control over the memory allocation and layout on all hierarchy levels, exposes deep copies between levels and does not assume a certain distribution of the memory over the various hierarchy levels. Specifically, it does not provide distributed data types that are intertwined with the parallelization scheme itself.

\textbf{Performance Portability} A model provides \emph{performance portability} if for a given \emph{single source} \emph{sustainable} and \emph{maintainable} implementation of an algorithm the hardware utilization on various systems is the same within reasonable margins, taking into account the limitations of the specific hardware. \emph{Performance portability} does not require optimum utilization of the hardware.














    

\section{Related Work}

In the following we briefly discuss other libraries targeting the portable parallel task execution within nodes.
Some of them require language extensions, others advertise performance portability across a multitude of devices.
However, none of these libraries can provide full control over the possibly diverse underlying hardware while being only minimally invasive.
Furthermore, many of the libraries do not satisfy the requirement for full single-source (C++) support.

\textbf{\cuda}\cite{CUDAPG}
is a parallel computing platform and programming model developed by \nvidia.
The user is bound to the usage of \nvidia GPUs. \cuda is not \emph{open source} and does not provide for \emph{Sustainability}, \emph{heterogeneity}, \emph{maintainability} and \emph{testability}. For \cuda enabled hardware it provides for \emph{optimizability}.

\textbf{PGI CUDA-X86}\footnote{\url{https://www.pgroup.com/resources/cuda-x86.htm}}
is a compiler technology that allows to generate x86-64 binary code from \cuda C/C++ applications using the \cuda runtime API but does not support the \cuda driver API. Compared to \cuda it allows for \emph{heterogeneity}, \emph{maintainability} and \emph{testability}, but it currently falls behind in adapting to the latest \cuda features, thus has limited support for \emph{sustainability}. As it does not provide for control of optimzations for X86 architectures, it lacks \emph{optimizability}.

\textbf{GPU Ocelot}\cite{kerr2011gpu}
is an \emph{open source} dynamic JIT compilation framework based on llvm which allows to execute native \cuda binaries 
by dynamically translating the \nvidia PTX virtual instruction set architecture to other instruction sets.
It supports \nvidia and \amd GPUs as well as multicore CPUs via a PTX to LLVM translator.
The project is not in active development anymore and only supports PTX up to version 3.1 while the current version is 4.2. Thus, it is in many respects similar to \textbf{PGI CUDA-X86}.

\textbf{\openmp}\footnote{\url{http://openmp.org//}}
is an \emph{open} specification for vendor agnostic shared memory parallelization which allows to easily parallelize existing sequential
C/C++/Fortran code in an incremental manner by adding annotations (pragmas in C/C++) to loops or regions.
Up to version 4.5 there is no way to allocate device memory that is persistent between kernel calls in different methods because
it is not possible to create a device data region spanning both functions in the general case.
Currently \openmp does not allow for controlling the hierarchical memory as its main assumption is a shared memory pool for all threads. Therefore, the block shared memory on \cuda devices cannot be explicitly utilized and both \emph{heterogeneity} and \emph{optimizability} are not provided for.
    
\textbf{OpenACC}\footnote{\url{http://www.openacc-standard.org/}}
is an \emph{open} pragma based programming standard for heterogeneous computing which is very similar to \openmp and provides annotations for parallel execution and data movement as well 
as run-time functions for accelerator and device management.
It allows for limited access to \cuda block shared memory but does not support dynamic allocation of memory in kernel code. It thus does not provide for \emph{optimizability} and in a practical sense, due to the very limited number of implementations, \emph{heterogeneity} and \emph{sustainability}.
    
\textbf{OpenCL}\footnote{\url{https://www.khronos.org/opencl/}}
is an \emph{open} programming framework for heterogeneous platforms. It supports \emph{heterogeneity} as it can utilize CPUs and GPUs of nearly all vendors.
Versions prior to 2.1 (released in March 2015) did only support a C-like kernel language.
Version 2.1 introduced a subset of C++14, but there are still no compilers available.
\opencl thus does not support \emph{single source} programming.
Furthermore, it does not allow for dynamic allocation of memory in kernel code and thus does not fully  support \emph{optimizability}.

\textbf{SYCL}\footnote{\url{https://www.khronos.org/sycl/}}
is an \emph{open} cross-platform abstraction layer based on \opencl and thus shares most deficiencies with \opencl, however it in principle would allow for \emph{optimizability}.
In contrast to \opencl it allows for \emph{single source} \emph{heterogeneous} programs, but as of now there is no usable free compiler implementation available that has good support for multiple accelerator devices, thus it currently lacks \emph{sustainability}.

\textbf{C++ AMP}\footnote{\url{https://msdn.microsoft.com/en-us/library/hh265136.aspx}}
is an \emph{open} specification from Microsoft which is implemented on top of DirectX 11 and thus currently limited in terms of \emph{heterogeneity}, \emph{sustainability} and \emph{testability}.
It is a language extension requiring compiler support that allows to annotate C++ code that then can be run on multiple accelerators. It lacks full control of parallel execution and memory hierarchy and thus falls short of supporting \emph{optimizability}. Due to restrictions on data types that provide for portability (see e.g. \texttt{concurrency::array}) it is not \emph{data structure agnostic}.

\textbf{KOKKOS}\footnote{\url{https://github.com/kokkos}}
is an \emph{open source} abstract interface for portable, high-performance shared memory-programming and in many ways similar to \alpaka.
However, kernel arguments have to be stored in members of the function object coupling algorithm and data together. It thus is not \emph{data structure agnostic} and in this sense limited in its \emph{optimizability}


\textbf{Thrust}\cite{bell2012thrust}
is an \emph{open source} parallel algorithms library resembling the C++ Standard Template Library (STL) which is available for \cuda, Thread Building Blocks\footnote{\url{https://www.threadingbuildingblocks.org/}} and \openmp back-ends at make-time. Its container objects are tightly coupled with the parallelization strategy, therefore \thrust is not \emph{data structure agnostic}. \thrust aims at hiding the memory hierarchy and is limited in expressing parallel execution, thus it cannot achieve full \emph{optimizability}.

Table~\ref{tab:distinction} provides a summary of all related work and a comparison to \alpaka.

\begin{table*}[!htbp]
  \scriptsize
  \caption{Properties of intra-node parallelization frameworks and their ability to solve the problems in porting high-performance HPC codes. \cmark : yes / fully solved, \omark : partially solved, \xmark : no / not solved}

	\begin{center}
          \begin{tabular}{ | m{2.5cm} | m{1cm} | m{1cm} | m{1cm} | m{1cm} | m{1.1cm} | m{1.4cm} | m{1cm} | m{1.3cm} | m{1.5cm} |}
	    \hline
	    Model & Openness & Single\newline Source & Sustain-ability & Hetero-geneity & Maintain-ability & Testability & Optimiz-ability & Data\newline structure\newline agnostic \\ \hline \hline
	    NVIDIA CUDA	     & \nosupp  & \issupp & \nosupp & \nosupp & \nosupp & \nosupp & \lisupp & \issupp \\ \hline
	    PGI CUDA-x86     & \nosupp	& \issupp & \lisupp & \issupp & \issupp & \issupp & \nosupp & \issupp \\ \hline
	    GPU Ocelot	     & \issupp  & \issupp & \lisupp & \issupp & \issupp & \issupp & \nosupp & \issupp \\ \hline
	    OpenMP	         & \issupp  & \issupp & \issupp & \lisupp & \lisupp & \issupp & \nosupp & \issupp \\ \hline
	    OpenACC	         & \issupp  & \issupp & \lisupp & \lisupp & \issupp & \issupp & \nosupp & \issupp \\ \hline
	    OpenCL	         & \issupp  & \lisupp & \issupp & \issupp & \issupp & \issupp & \nosupp & \issupp \\ \hline
	    SYCL	         & \issupp  & \issupp & \lisupp & \issupp & \issupp & \lisupp & \lisupp & \issupp \\ \hline
	    C++AMP	         & \issupp  & \issupp & \lisupp & \lisupp & \issupp & \lisupp & \nosupp & \lisupp \\ \hline
	    KOKKOS	         & \issupp  & \issupp & \issupp & \issupp & \issupp & \issupp & \nosupp & \lisupp \\ \hline
	    Thrust	         & \issupp  & \issupp & \issupp & \issupp & \issupp & \issupp & \nosupp & \nosupp \\ \hline
	    \textbf{\alpaka} & \issupp  & \issupp & \issupp & \issupp & \issupp & \issupp & \issupp & \issupp \\ \hline
	    \end{tabular}
		\label{tab:distinction}
	\end{center}
        \vspace{-2em}        
\end{table*}

\section{Introduction to Alpaka}
This section serves as an introduction to \alpaka.
It first explains the conceptual ideas behind \alpaka, then provides an overview of the hardware abstraction model of \alpaka as well as how the model is mapped to real devices.
Lastly, the \alpaka programming API is described.

\subsection{Conceptual Overview}
\label{subsec:design}
\alpaka provides a single abstract C++ interface to describe parallel execution across multiple levels of the parallelization hierarchy on a single compute node. Each level of the \alpaka parallelization hierarchy is unrestricted in its dimensionality.
In addition, \alpaka uses the offloading model, which separates the host from the accelerator device.

In order to execute \alpaka code on different hardware the interface is currently implemented using various parallelism models such as \openmp, \cuda, C++ threads and boost fibers. \alpaka interface implementations, called back-ends, are not limited to these choices and will in the future be extended by e.g. Thread Building Blocks. By design, new back-ends can be added to \alpaka.
Thus, \alpaka allows for mixing parallelization models in a single source program, thereby enabling the user to choose the implementation that is best suited for a given choice of hardware and algorithm.
It even enables running multiple of the same or different back-end instances simultaneously, e.g. to utilize all cores on a device as well as all accelerators concurrently.

The \alpaka library is based on the C++11 standard without any language extensions and makes extensive usage of C++ template meta-programming.
Algorithms are written in-line with single source code and are called kernels which can be compiled to multiple platform back-ends by simply selecting the appropriate back-end. The actual back-ends that execute an algorithm can, thus, be selected at configure-, compile- or run-time, making it possible to run an algorithm on multiple back-ends in one binary at the same time.

\alpaka does not automatically optimize data accesses or data transfers between devices.
Data are stored in simple buffers with support for copies between devices and access to memory is completely data structure agnostic. Thus, the user needs to take care of distribution of data between devices.

\alpaka does neither automatically decompose the algorithmic execution domain and data domain, nor does it assume any default or implicit states such as default device, current device, default stream, implicit built-in variables and functions.

\subsection{Model of Parallel Abstraction}
\label{subsec:model}
\alpaka abstracts data parallelism following the redundant hierarchical parallelism model~\cite{rocki2014future}, thereby enabling the developer to explicitly take the hierarchy of processing units, their data parallel features and corresponding memory regions into account.
The \alpaka abstraction of parallelization is influenced by and based on the groundbreaking \cuda and \opencl abstractions\footnote{Both, \cuda and \opencl are industry standards for accelerator programming.} of a multidimensional grid of threads with additional hierarchy levels in between. Furthermore, it is amended with additional vectorization capabilities.

\begin{figure}[t]
  \centerline
      {\resizebox{0.5\textwidth}{!}{\includegraphics{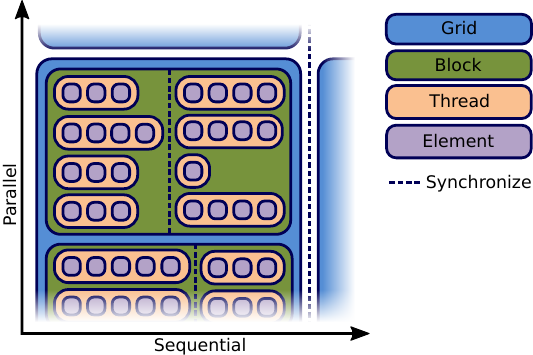}}}
      \caption{The \alpaka parallelization hierarchy consists of a grid of blocks, where each block consists of threads and each thread processes multiple elements.
        Both threads and grids are able to synchronize.
      }
      \label{fig:model}
\end{figure}

The four main hierarchies introduced by \alpaka are called \emph{grid}, \emph{block}, \emph{thread} and \emph{element} level, shown in Figure~\ref{fig:model} together with their respective parallelization and synchronization features as discussed below.

\begin{figure}[th]
  \centerline
      {\resizebox{0.5\textwidth}{!}{\includegraphics{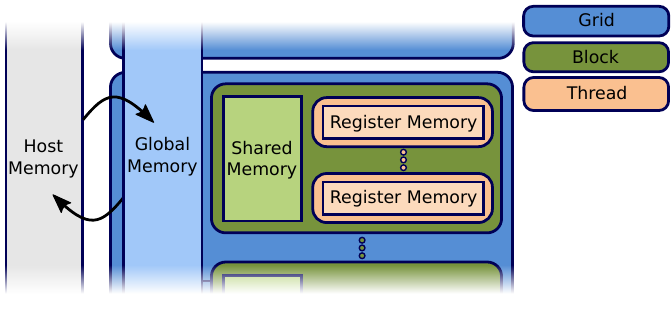}}}
      \caption{The memory hierarchy of the \alpaka abstraction model.
        Threads have exclusive access to fast register memory.
        All threads in a block can access the same shared memory.
        All blocks in a grid can access the same global memory.
      }
      \label{fig:memory}
      \vspace{-0.5em}
\end{figure}

Each parallelization level corresponds to a particular memory level (Figure \ref{fig:memory}): global memory~(grid), shared memory~(block) and register memory~(thread).

The \alpaka model enables to separate the parallelization strategy from the algorithm.
The algorithm is described by \emph{kernel} functions that are executed by threads.
A \emph{kernel} is the common set of instructions executed by all threads on a grid.

The parallelization strategy is described by the accelerator and the work division (See Section~\ref{subsec:mapping} and \ref{subsec:interface}).
An accelerator defines the acceleration strategy by a mapping of the parallelization levels to the hardware.
The device is the actual hardware onto which these levels are mapped.
\\
\subsubsection{\textbf{Grid}}
A grid is an n-dimensional set of blocks with a usually large global memory accessible by all threads in all blocks.
Grids are independent of each other and can thus be executed either sequentially or in parallel. Grids can be synchronized to each other via explicit synchronization evoked in the code.

\subsubsection{\textbf{Block}}
A block is an n-dimensional set of threads with a high bandwidth, low latency but small amount of shared memory.
All blocks on a grid are independent of each other and can thus be executed either sequentially or in parallel. Blocks cannot be synchronized to each other.
The shared memory can only be accessed explicitly by threads within the same block and gets discarded after the complete block has finished its calculation.

\subsubsection{\textbf{Thread}}
A thread represents the execution of a sequence of instructions.
All threads in a block are independent of each other and can thus be executed either sequentially or in parallel. Threads can be synchronized to each other via explicit synchronization evoked in the code.
Threads can by default always access their private registers, the shared memory of the block and the global memory\footnote{However, \alpaka allows for atomic operations that serialize thread access to global memory.}.
All variables within the default scope of a kernel are stored within register memory and are not shared between threads.
Shared and global memory can be allocated statically or at runtime before the kernel start.

\subsubsection{\textbf{Element}}
The element level represents an n-dimensional set of elements and unifies the data parallel capabilities of modern hardware architectures e.g. vectorization on thread level.
This is necessary as current compilers do not support automatic vectorization of basic, non trivial loops containing control flow statements~(e.g. if, else, for) or non-trivial memory operations.
Furthermore, vectorization intrinsics as they are available in intrin.h, arm\_neo.h, altivec.h are not portable across varying back-ends.
\alpaka therefore currently relies on compiler recognition of vectorizable code parts. Code is refactored in such a way that it includes primitive inner loops over a fixed number of elements.

The user is free to sequentially loop over the elements or to utilize vectorization where a single instruction is applied to multiple data elements in parallel e.g. by utilizing SIMD vector registers.
Processing multiple elements per thread on some architectures may enhance caching.

\subsection{Mapping of Abstraction to Hardware}
\label{subsec:mapping}
\alpaka clearly separates its parallelization abstraction from the specific hardware capabilities by an explicit mapping of the parallelization levels to the hardware.
A major point of the hierarchical parallelism abstraction is to ignore specific unsupported levels of the model and utilize only the ones supported on a particular device. Mapping is left to the implementation of the accelerator.

This allows for variable mappings as shown in the examples below and, therefore, an optimum usage of the underlying compute and memory capabilities---albeit with two minor limitations: The grid level is always mapped to the whole device being in consideration and the kernel scheduler can always execute multiple kernel grids from multiple streams in parallel by statically or dynamically subdividing the available resources.

\begin{figure}[th]
  \centerline
      {\resizebox{0.5\textwidth}{!}{\includegraphics{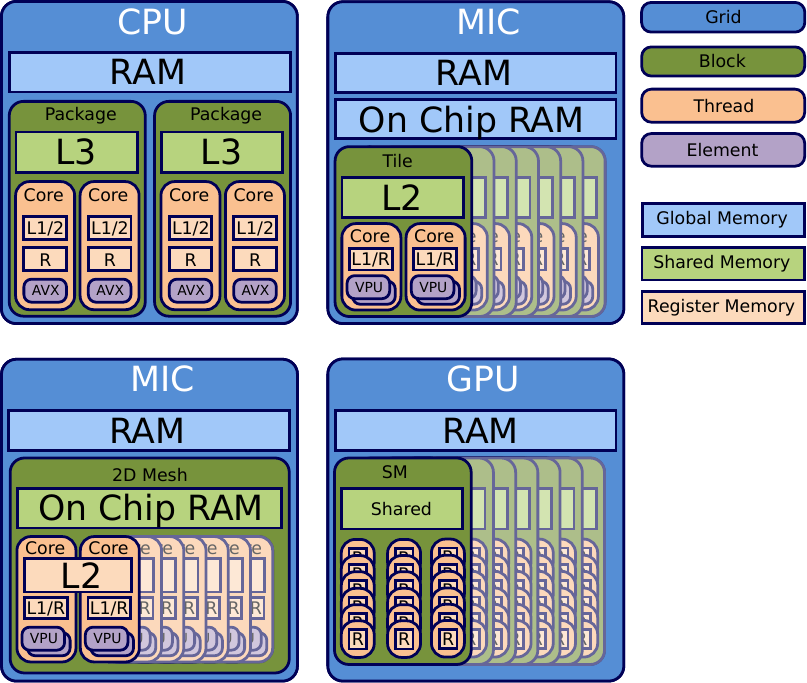}}}
      \caption{Possible mapping of blocks, threads and elements to a MIC, a CPU and a GPU device. The mapping can skip individual levels when they are not beneficial on a particular device.}
      \label{fig:mapping}
      \vspace{-0.5em}
\end{figure}

Figure~\ref{fig:mapping} shows a mapping of the \alpaka abstraction model onto a CPU, a many integrated cores device~(MIC) and a GPU architecture.
For the MIC architecture a second mapping is shown, which spans a block over all cores to increase the shared memory.
CPU and MIC process multiple elements per thread and benefit from their vectorization units, while a GPU thread processes only a small amount of elements.

\noindent Finally, the user needs to decide which back-end to use for which device.
It can be selected from the set of predefined accelerators or the user can write its own accelerator implementation.
The set of predefined accelerator mappings are listed in Table~\ref{tab:accelerators}.
\begin{table}[!htbp]
  \scriptsize
  \caption{Predefined accelerators with: problem size(N), threads per block(B), elements per thread(V).}
  \begin{center}
    \begin{tabular}{ | l | l | c | c | c | c |}
      \hline
      Arch & Acc & Grid & Block & Thread & Element\\
      \hline
      \hline
      \textbf{GPU} & \cuda & $1$ & $N/(B \cdot V)$ & $B$ & $V$\\
      \hline
      \textbf{CPU} & \openmp block & $1$ & $N/V$ & $1$ & $V$\\
      & \openmp thread & $1$ & $N/(B \cdot V)$ & $B$ & $V$\\
      & C++11 thread & $1$ & $N/(B \cdot V)$ & $B$ & $V$\\
      & Sequential & $1$ & $N/V$ & $1$ & $V$\\
      \hline
      \textbf{MIC} & \openmp block & $1$ & $N/V$ & $1$ & $V$\\
      & \openmp thread & $1$ & $N/(B \cdot V)$ & $B$ & $V$\\
      \hline
    \end{tabular}

  \end{center}
  \label{tab:accelerators}
\end{table}
\subsection{Alpaka Programming Interface}
\label{subsec:interface}
In the following each part of the \alpaka interface is described briefly.
The provided listings assume that the \alpaka namespace is used.
Source code examples are provided to give a more detailed insight into \alpaka.

\hfill
\subsubsection{Kernel}
The kernel is the central unit in \alpaka that acts as the bridge between host and accelerator code through a C++ class or a C++ lambda (C++14 required).
The algorithm is described from the block down to the element level which removes the need for a nested loop structure like it is used in \openmp and  SYCL.
The kernel function object needs to implement the template \cpp{operator()} member function as it is shown in Listing~\ref{lst:kernel_skeleton}.
This member function is the first function called on the particular accelerator device.

\begin{figure}
\begin{minipage}{\linewidth}
\hfill
  \begin{lstlisting}[caption={A skeleton of an \alpaka kernel. The kernel needs to implement the operator() with prefix ALPAKA\_FN\_ACC, which takes at least the accelerator as parameter.}, label={lst:kernel_skeleton}, escapeinside={|}{|}]
struct Kernel {

  template <class T_Acc, class T>
  ALPAKA_FN_ACC void operator()(T_Acc acc,|\label{lst:kernel_skeleton:4}|
                                T data) const {

    /* Write kernel code here */

  }

};
  \end{lstlisting}
\end{minipage}
      \vspace{-2.5em}
\end{figure}

\noindent The code within the kernel is written in C++11 (restricted by the utilized back-end compilers) with additional calls to the \alpaka run-time API.
There exist no implicit built-in variables and functions like it is usual in \cuda or \opencl.
All information can be retrieved from the accelerator object~(Listing~\ref{lst:kernel_skeleton}~:~\ref{lst:kernel_skeleton:4}).
\subsubsection{Accelerator Executable Functions}
\alpaka defines the macros ALPAKA\_FN\_HOST, ALPAKA\_FN\_ACC and ALPAKA\_FN\_HOST\_ACC to define that functions are callable from host, from accelerator or from both host and accelerator device.
All functions called from accelerator code need to be prefixed by these macros.
%
\subsubsection{Work Division and Index Retrieval}
The work division defines the extent and dimensionality of each level of the \alpaka abstraction model.
Listing~\ref{lst:workdiv_host} shows the declaration of a two-dimensional work division on the host where the grid level has an extent of 128 blocks and the other levels have an extent of one.

\begin{figure}
\begin{minipage}{\linewidth}
\hfill
\begin{lstlisting}[caption={Declaration of a work division in the host code. The work division is defined in two dimensions for all levels. Element and block level have an extent of one, while the grid has an extent of 128.}, label={lst:workdiv_host}, escapeinside={|}{|}]
Vec<Dim2, size_t> elementsPerThread(1,1);
Vec<Dim2, size_t> threadsPerBlock(1,1);
Vec<Dim2, size_t> blocksPerGrid(8,16);

workdiv::WorkDivMembers<Dim2, size_t>
  (blocksPerGrid,
   threadsPerBlock,
   elementsPerThread);
\end{lstlisting}
\hfill
\end{minipage}
      \vspace{-1.5em}
\end{figure}

\noindent The work division declared in Listing~\ref{lst:workdiv_host} can be accessed within the kernel via the accelerator object.
Furthermore, there exist methods to map the index space between varying extents and dimensionalities.
Listing~\ref{lst:workdiv_dev} shows a kernel function that calculates the global linearized index of a thread with the help of \alpaka run-time functions.
\begin{figure}
\begin{minipage}{\linewidth}
\hfill
  \begin{lstlisting}[caption={Access of work division and thread index. Thread extent and index are mapped onto a one dimensional space to retrieve a linearized index.}, label={lst:workdiv_dev}, escapeinside={|}{|}]
ALPAKA_FN_ACC void operator()(T_Acc acc) const {

  // Retrieve the global n-dim thread index
  auto gTIdx = idx::getIdx<Grid, Threads>(acc);

  // Retrieve the n-dim thread extent
  auto gTExtent = workdiv::getWorkDiv<Grid, Threads>(acc);

  // Retrieve the global one dim thread index
  auto linIdx = core::mapIdx<1>(gTIdx,gTExtent);
}
  \end{lstlisting}
\hfill
\end{minipage}
      \vspace{-1.5em}
\end{figure}
\subsubsection{Memory}
\alpaka provides simple memory buffers that store the plain pointer to memory of the particular device and additional information like residing device, extent, pitch and dimension.
These buffers are uniform for all devices which allows for copying memory between different devices with respect to pitch and extents, see Listing~\ref{lst:buffer}.

\begin{figure}
\begin{minipage}{\linewidth}
\hfill
  \begin{lstlisting}[caption={Allocation of a two dimensional host and a device buffer with one hundred elements each. Moreover, the host buffer is copied to the device buffer.}, label={lst:buffer}, escapeinside={|}{|}]
// Dim, data and index type
using Dim  = dim::DimInt<2>;
using Data = std::uint32_t;
using Size = std::size_t;

// Declare extents of buffer
Vec<Dim, Size> extents(10,10);

// Declare host and device buffer
auto hostBuf = mem::buf::alloc<Data,Size>(host, extents);
auto devBuf  = mem::buf::alloc<Data,Size>(dev, extents);

// Copy host buffer to device buffer
mem::view::copy(stream, devBuf, hostBuf, extents);
  \end{lstlisting}
\hfill
\end{minipage}
      \vspace{-1.5em}
\end{figure}
\subsubsection{Streams}
A stream is the work queue of a particular device.
Operations in streams are always executed in-order: No operation in a stream will begin before all previously issued operations in the stream have completed.
Streams can be synchronous or asynchronous with respect to operations on the host.
If an operation is issued in a synchronous stream, the host thread will block until this operation is finished.
Asynchronous streams allow the host to resume computations while the accelerator is executing the operation.
%
\subsubsection{Kernel Execution}
A kernel will be executed by enqueuing a kernel executor into a stream of a particular device.
An executor binds an accelerator, a work division, a kernel and its parameters.
Streams are filled with those executors and \alpaka takes care that they will be executed in the specified way.
Listing~\ref{lst:kernel} shows a full host code example from the accelerator type definition up to the enqueuing of a kernel into a stream.

\begin{minipage}{\linewidth}
\hfill
\begin{lstlisting}[caption={Full example of a kernel execution.
      The kernel is executed with a work division of 256 blocks and 16 threads per block on a single accelerator (in this case the sequential CPU back-end is selected).}, label={lst:kernel}, escapeinside={|}{|}]
// Define the dimensionality of the task
using Dim  = dim::DimInt<1u>;
using Size = std::size_t;

// Define the accelerator and stream to use
using Acc    = acc::AccCpuSerial<Dim, Size>;
using Stream = stream::StreamCpuAsync;

// Select a device to execute on
auto devAcc = dev::DevMan<Acc>::getDevByIdx(0);

// Create a stream to enqueue the execution into
Stream stream(devAcc);

// Create a 1d work division with 256 blocks
// a 16 threads a 1 element
auto workDiv(workdiv::WorkDivMembers<Dim, Size>(256u, 16u, 1u);
// Create an instance of the kernel
Kernel kernel;
// Create the execution task
auto exec(exec::create<Acc>(workDiv, kernel/*, arguments ...*/);

// Enqueue the task into the stream
stream::enqueue(stream, exec);
  \end{lstlisting}
\hfill
\end{minipage}
      \vspace{-2em}

\section{Evaluation}
This section provides an evaluation of \alpaka on a variety of hardware platforms using various back-ends, see
Table~\ref{tab:hardware}.
All \cuda evaluations are compiled with \cuda 7.0 and all CPU evaluations with gcc 4.9.2.
Source codes denoted as \emph{native} are not wrapped by \alpaka, but contain pure \cuda or \openmp code.

The evaluation was performed in five stages:
First, the PTX and assembler code generated during compilation of an \alpaka DAXPY program is compared to the respective native versions.
Then, the overhead of two \naive \alpaka DGEMM kernels with respect to their native versions is measured.
As a next step, it is investigated what happens to performance portability when the \naive \alpaka DGEMM kernels are mapped to an inappropriate back-end followed by the description of a single source DGEMM kernel and how it can obtain performance portability with the help of \alpaka.
Finally, the applicability of \alpaka to real world applications was evaluated using HASEonGPU~\cite{haseongpu}.

\begin{table*}[!htbp]
  \scriptsize
  \begin{center}
    \caption{List of utilized accelerator hardware for evaluation. Clock frequencies which are encapsulated in braces denote the turbo frequency of the particular architecture. Often turbo can
only be utilized when not all cores of a device are busy.}
    \begin{tabular}{ |l || c | c | c |  c | c |}
      \hline
      \textbf{Vendor}                       & AMD            & \intel               & \intel              \ignore{& IBM}           & \nvidia          & \nvidia\\
      \hline
      \textbf{Architecture}                 & Opteron 6276   & Xeon E5-2609        & Xeon E5-2630v3       \ignore{& Power8 8247-42L}      & K20 GK110       & K80 GK210\\
      \hline
      \textbf{Number of devices}            & 4              & 2                   & 2                    \ignore{& 2}              & 1         & 2\\
      \hline
      \textbf{Number of cores per device}   & 16             & 4                   & 8 (16 hyper-threads) \ignore{& 10 (80)}       & 2496            & 2x2496\\
      \hline
      \textbf{Clock frequency}              & 2.3 (3.2) GHz  & 2.4 GHz             &  2.4 (3.2) GHz       \ignore{& 2.1 (3.69) GHz}  & 0.56 (0.88) GHz\\
      \hline
      \textbf{Release date}                 &   Q4/2011      &   Q1/2012           &   Q3/2014            \ignore{&  Q1/2014}      & Q4/2012        & Q4/2014\\
      \hline
      \textbf{Th. double peak performance}  & 480 GFLOPS     & 150 GFLOPS          &  540 GFLOPS          \ignore{& 560 GFLOPS}    & 1170 GFLOPS     & 2x1450 GFLOPS\\
      \hline
    \end{tabular}
    \label{tab:hardware}
  \end{center}
        \vspace{-2em}
\end{table*}

\subsection{Conceptual Comparison}
This section compares the \alpaka implementation of the \emph{generalized vector addition algorithm} in double precision~(DAXPY) to a sequential C++ and \cuda implementation on the source code and assembler level.

DAXPY computes $\boldsymbol{Y} \leftarrow \alpha \boldsymbol{X} + \boldsymbol{Y}$ where $\boldsymbol{X}$ and $\boldsymbol{Y}$ are vectors while $\alpha$ is a scalar.
DAXPY was selected as a trivial example on the one hand to showcase the non obfuscation abstraction and on the other hand the zero overhead abstraction of \alpaka.
The source code of the various DAXPY implementations are available in our \github repository\footnote{\url{https://github.com/BenjaminW3/vecadd}}.

From a developers point of view, the source codes are very similar to each other.
However, \alpaka related changes are necessary:
\alpaka adds the additional accelerator template argument together with the according function argument to the kernel function call.
Each function that should be called from a accelerator needs to be annotated with the \alpaka specific macro \cpp{ALPAKA\_FN\_ACC}.
Furthermore, \alpaka replaces the for loop index calculations by an \alpaka equivalent that calculates the correct index for each thread.

%

\begin{figure}[tb]
  \centerline
      {\resizebox{0.5\textwidth}{!}{\includegraphics{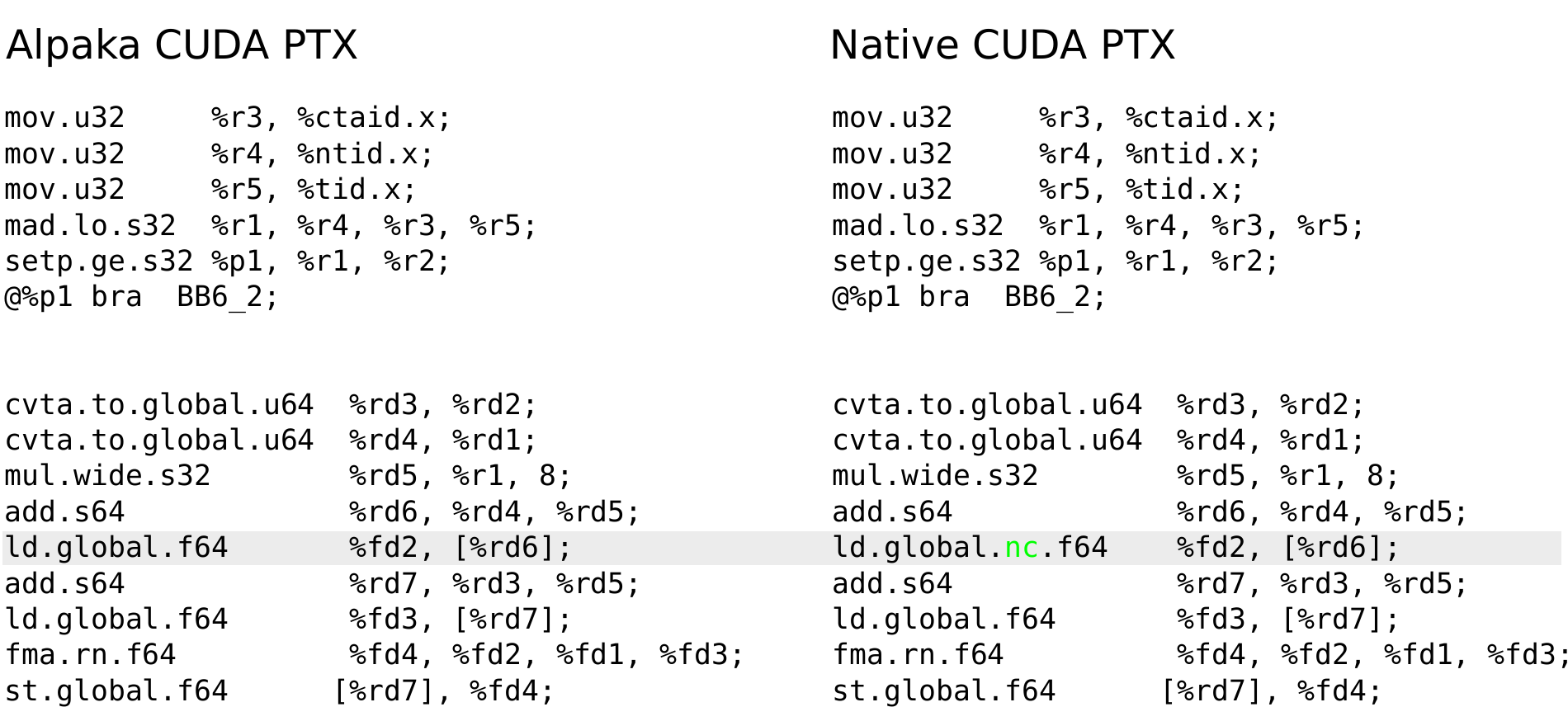}}}
          \caption{Snippet of the PTX code of \alpaka and \cuda kernels. The PTX code is the same up to the line where the \cuda PTX uses non coherent texture cache. This cache allows for access with higher bandwidth and lower latency.}
          \label{fig:ptx}
          \vspace{-1em}
\end{figure}

Figure~\ref{fig:ptx} shows snippets of the PTX codes of the compiled \alpaka and \cuda implementations.

Comparing the generated PTX code leads to the result that these codes are identical up to two additional but unused function parameters in the \alpaka variant as well as different internal variable names and the use of non coherent texture cache once.
It can be seen that modern compilers are able to remove all the meta-programming abstraction introduced by \alpaka.
This perfectly demonstrates the zero overhead abstraction of the \alpaka interface regarding the \cuda interface.

The assembler code of the native C++ implementation does not perfectly fit the \alpaka assembler since only the native implementation has been vectorized to use the packed double precision SSE2 instruction \cpp{movupd}, \cpp{mulpd} and \cpp{addpd} instead of the single value versions \cpp{movsd}, \cpp{mulsd} and \cpp{addsd}.
However, by looping over the additional element level of the \alpaka abstraction model which has a constant size, the compiler recognizes the iteration independent looping pattern and optimizes this by using \simd instructions to process multiple consecutive iterations together.

\subsection{Performance}
As a next step, the performance characteristics of the \cuda and \openmp \alpaka back-ends are evaluated.
First, an algorithm is implemented for both \alpaka and the particular native API to show the pure \alpaka overhead in numbers.
Then, the native \alpaka kernel is mapped to the non-native back-end to show that \alpaka is not na\"{i}vely performance portable.
Afterwards, an enhanced  single source \alpaka kernel is introduced and mapped to various architectures and it is shown that it can match the performance of the native implementations when using the appropriate \alpaka back-ends.

For comparison the \emph{double generalized matrix-matrix-multiplication} (DGEMM) has been selected as a compute bound problem.
DGEMM computes $\boldsymbol{C} \leftarrow \alpha \boldsymbol{A} \boldsymbol{B} + \beta \boldsymbol{C}$ where $\boldsymbol{C}$, $\boldsymbol{A}$ and $\boldsymbol{B}$ are matrices $A = (a_{i,j})$, $B = (b_{i,j})$ while $\alpha$ and $\beta$ are scalars.
DGEMM implementations can utilize all levels of parallelism on GPUs and CPUs ranging from blocking over shared memory to vectorization, providing a perfect showcase of these techniques within the \alpaka library.
All DGEMM implementations are available in our \github repository~\footnote{\url{https://github.com/BenjaminW3/matmul}}.

All input matrices are dense and always have square extents to minimize bias towards implementations preferring column- or row-major layout.
Initially, the matrices are filled with random values in the range~[0.0,~10.0].
The matrices are mapped to 1D memory buffers with \alpaka aligning rows to optimum memory boundaries.
Measurements do not include times for allocating the matrices on the host, filling them, a possible data transfer between the processor and a co-processor as well as device and stream initialization.

\hfill
\subsubsection{Zero Overhead Abstraction}
\alpaka does not add additional overhead to the algorithm execution time.
In order to show this zero overhead, native \cuda and \openmp~2 kernels were translated one-to-one to \alpaka kernels.

\begin{figure}[tb]
  \centerline
      {\resizebox{0.5\textwidth}{!}{\includegraphics{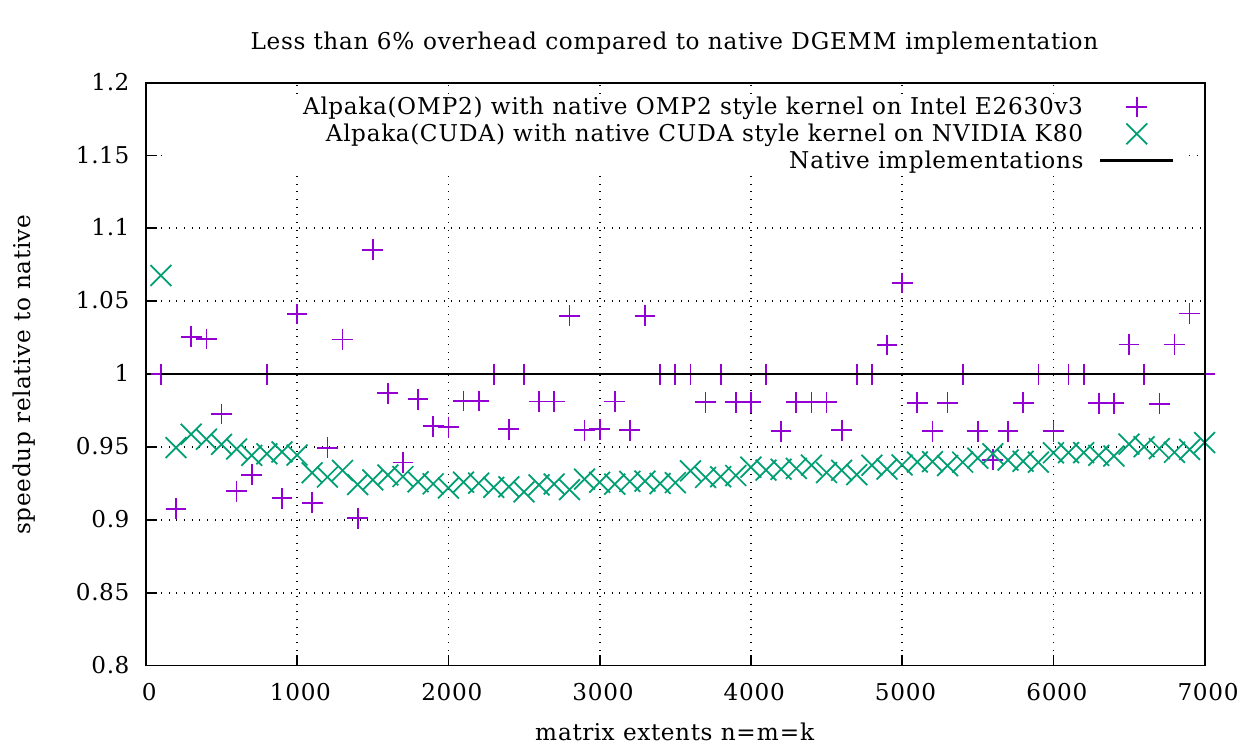}}}
      \caption{The native \alpaka kernels were mapped to their corresponding native back-ends and compared to the native implementations.
        Both kernels show a relative overhead of less than 6\% which is well below run-to-run variation. This proves the zero overhead abstraction of \alpaka.}
  \label{fig:performance_overhead}
\end{figure}

The \cuda kernels use a block parallelized tiling algorithm based on the \cuda programming guide (\cite{CUDAPG}, Sec.~3.2.3) and were executed on a compute node with a \nvidia K20 GK210.
The \openmp kernels use a standard DGEMM algorithm with nested for loops and were executed on a compute node with two \intel E5-2630v3 CPUs.
The kernels were executed with an increasing matrix size and their execution time was measured.
Figure~\ref{fig:performance_overhead} shows the speed of the \alpaka kernels mapped to the corresponding back-end relative to their native implementations.

The native \cuda \alpaka kernel provides more than 94\% relative performance for almost all matrix sizes, which is an overhead of 6\% or less.
After a deep inspection of the compiled PTX code it turned out that this overhead results from move and forward operators translated to copies.
These operators are used for grid index calculations within an \alpaka kernel.
Furthermore, a small number of additional \cuda runtime calls by the alpaka \cuda back-end are necessary.
The native \openmp \alpaka kernel provides an average relative performance of 100\%.

\begin{figure}[tb]
  \centerline
      {\resizebox{0.5\textwidth}{!}{\includegraphics{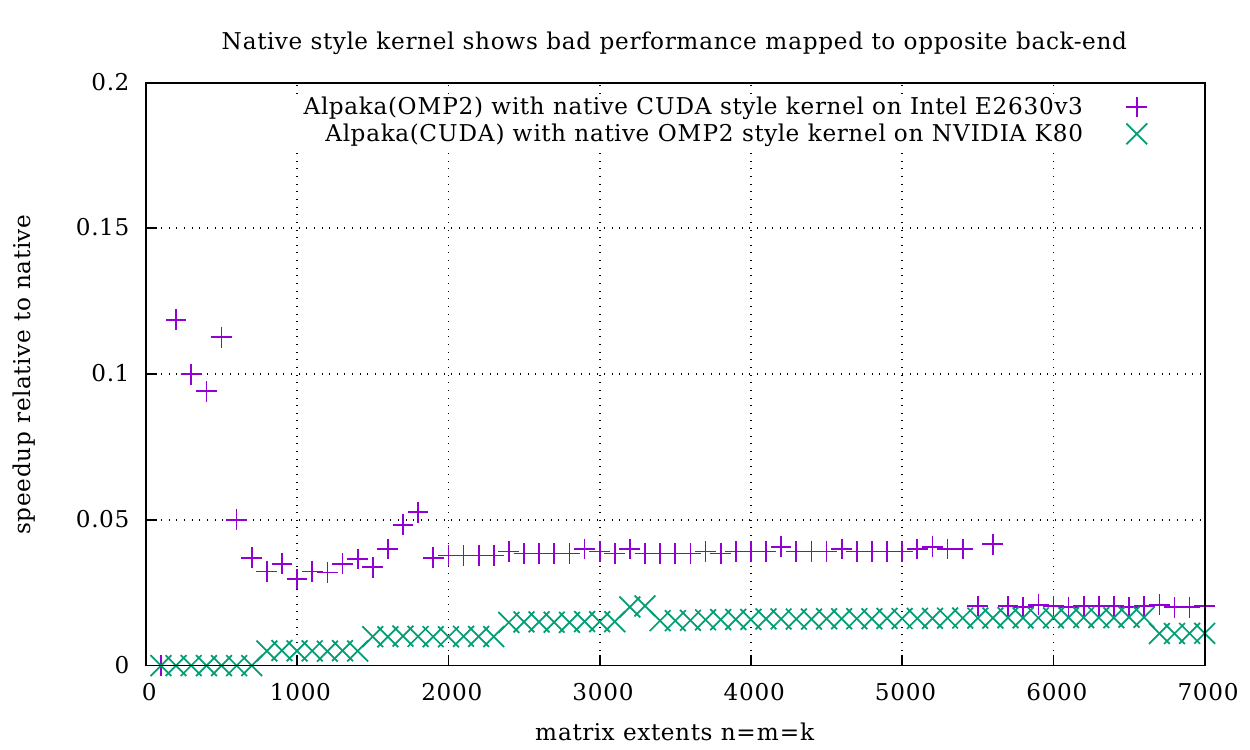}}}
      \caption{The native \alpaka kernels with swapped back-ends leads to poor performance.
        \alpaka does not guarantee performance portability when data access, work division and cache hierarchies are not considered.}
  \label{fig:performance_overhead_bad}
\end{figure}

One-to-one translation of a particular algorithm to an \alpaka kernel demonstrates a minimal amount of overhead compared to the native implementation on the same architecture.
However, \alpaka does not guarantee that such a kernel will also show the same run-time characteristics when it is mapped onto another back-end, as it neither provides optimized nor architecture dependent data access and work division automatically.
Figure~\ref{fig:performance_overhead_bad} shows the performance of the previously used kernels when their back-ends are swapped relative to the native implementation\footnote{In this case the triple nested loop is compiled using the CUDA back-end, while the tiled shared-memory version is mapped to OpenMP.}.

The \alpaka kernels originally translated from the opposite back-end do not perform well.
There are at least two reasons why these mappings are not performance portable.
First, the back-ends require completely different data access patterns to achieve optimum data access performance e.g. strided data access in \cuda.
Second, the amount of data a single thread can process effectively is different because of different cache sizes and hierarchies and varying optimal work divisions.

Nevertheless, it is possible to write competitive code for each back-end.
Both, the \nvidia~\cuda~(nvcc) and the gcc compiler remove all the abstraction layers introduced by \alpaka.

A \naive port of a kernel to an architecture it was not meant to be executed on will almost always lead to poor performance.
Thus, providing a single, performance portable kernel is not trivial.
The following section shows that \alpaka is able to provide performance for various back-ends with a single source kernel.

\hfill
\subsubsection{Single Source Kernel~/~Performance}
It is possible to write a single source kernel that performs well on all tested \alpaka back-ends without a drop in performance compared to the native implementations.
In order to reach this performance, the developer needs to abstract the access to data, optimize the work division, and consider cache hierarchies.
The single source \alpaka DGEMM kernel implements a tiling matrix-matrix multiplication algorithm and considers the architecture cache sizes by adapting the number of elements processed per thread or block and the size of the shared memory to provide minimal access latency.
A lot of processor architectures benefit from the \alpaka element level parallelism when calculating multiple elements in parallel in the vector processing unit.

\begin{figure}[tb]
  \centerline
      {\resizebox{0.5\textwidth}{!}{\includegraphics{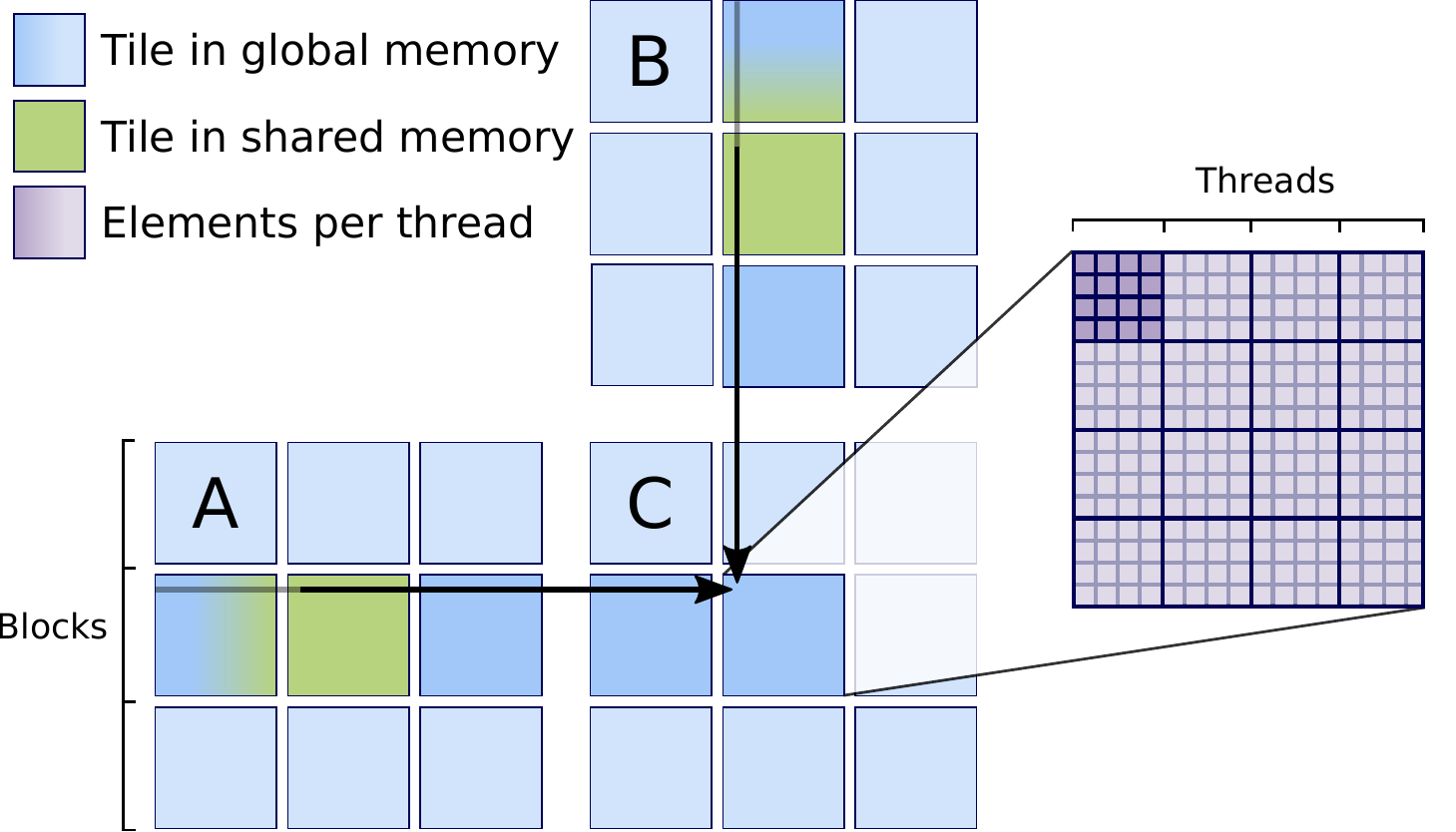}}}
      \caption{An \alpaka optimized hierarchically tiled matrix-matrix multiplication algorithm with multiple elements per thread.
        A block loads tiles of the A and B matrix into its shared memory to increase memory reuse.
        A thread can calculate multiple elements by using the vector processing unit of its particular back-end.}
  \label{fig:tiling_kernel}
\end{figure}

Figure~\ref{fig:tiling_kernel} provides a brief description of the hierarchical tiling algorithm.
A block calculates the result of a tile in matrix C.
Each thread in this block loads a set of elements of matrices A and B into shared memory to increase memory reuse.
It then calculates the partial results of its set of elements before the block continues with the next tiles of A and B.

\begin{figure}[bt]
  \centerline
      {\resizebox{0.5\textwidth}{!}{\includegraphics{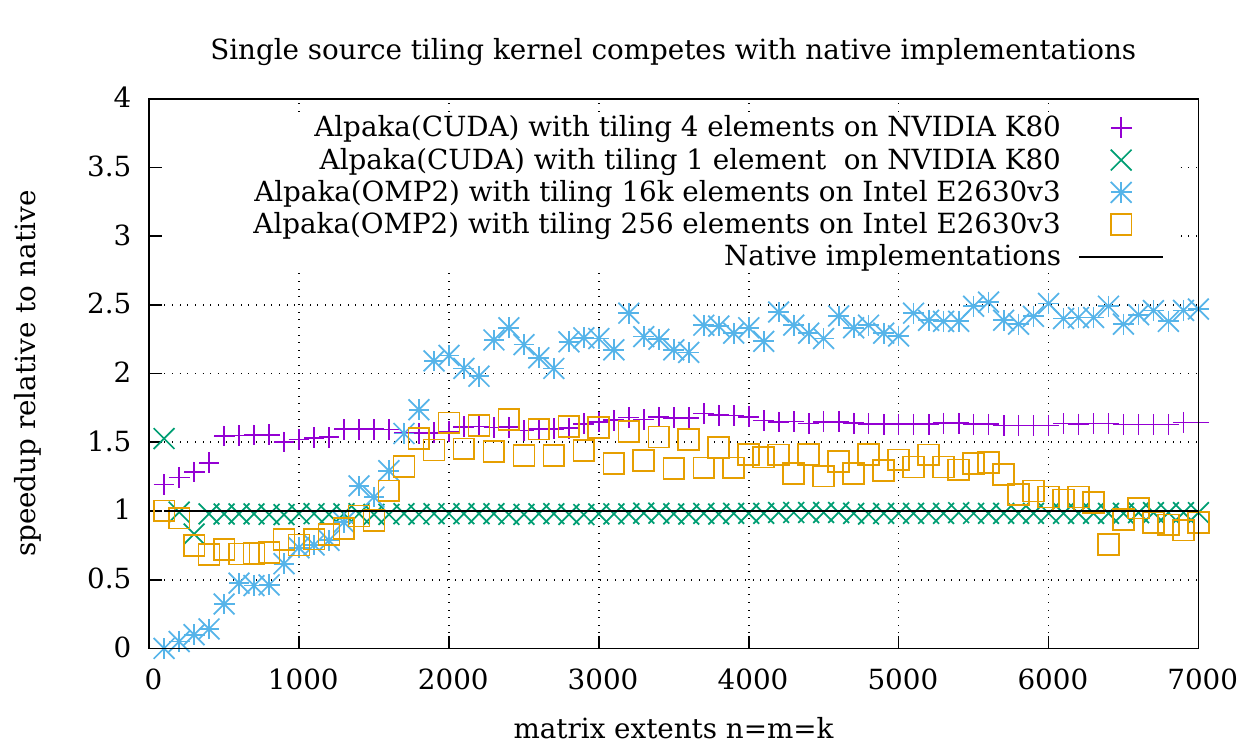}}}
      \caption{The \alpaka single source DGEMM kernel implements a hierarchical tiling matrix-matrix multiplication algorithm.
        This kernel can compete with and even outperform the original native implementations on all tested back-ends.}
  \label{fig:relative_performance_tiling_kernel}
\end{figure}

Figure~\ref{fig:relative_performance_tiling_kernel} shows the performance of the tiling kernel mapped to the \cuda and \openmp back-ends relative to the original native implementations. No performance loss compared to native implementations is observed but instead performance gain in the majority of cases is seen.
This is due to the more descriptive nature of the \alpaka kernel which enables even further optimizations by the back-end compilers.

It is clear that there exist even more optimized versions of the algorithm, e.g. in mathematical libraries such as \cublas, which is fine tuned for different compute-capabilities of \nvidia GPUs, or MKL, which is an optimized \openmp kernel library.
These  provide higher peak performance than \alpaka, but may require additional setup (\cublas data transfers) or include implicit---and maybe unwanted---data migration between the host and the device.
Nevertheless, if it should be necessary to use one of these optimized algorithms it is possible to use them with \alpaka as well by utilizing template specialization within \alpaka kernels.

\hfill
\subsubsection{Performance Portability}

\begin{figure}[tb]
  \centerline
      {\resizebox{0.5\textwidth}{!}{\includegraphics{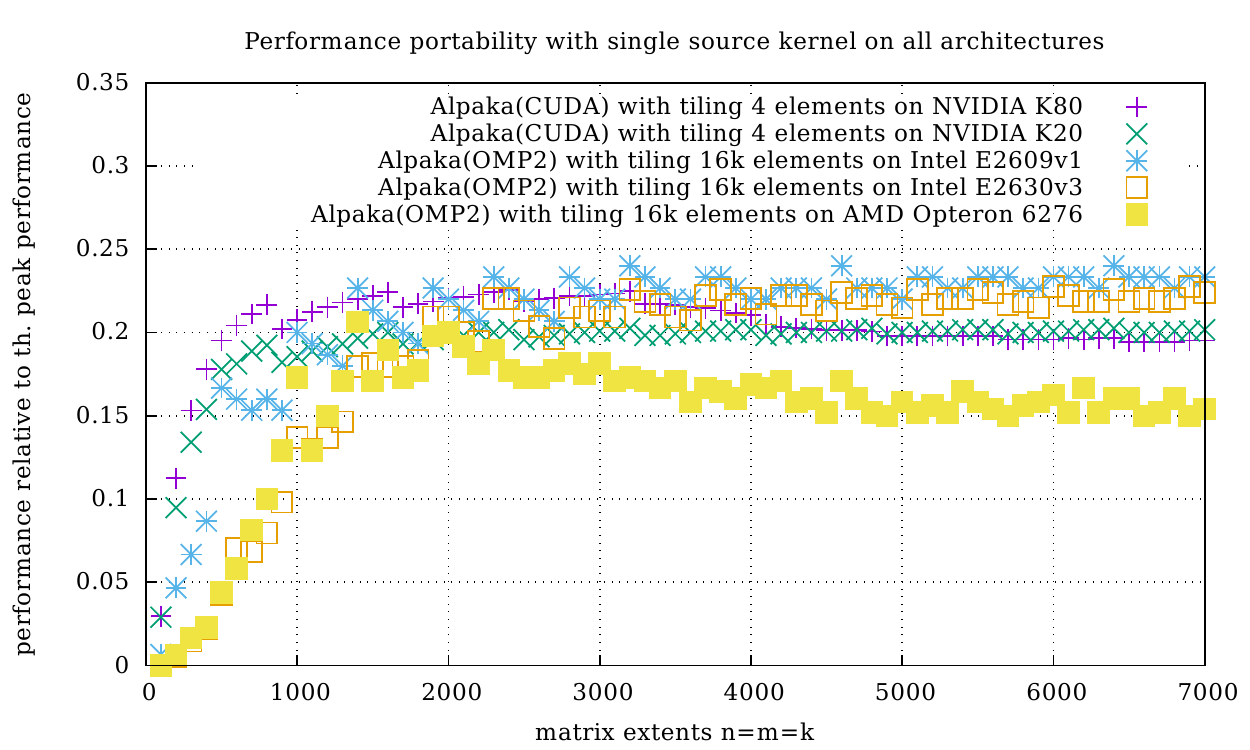}}}
      \caption{Performance of the \alpaka kernel executed on various architectures relative to the theoretical peak performance of the corresponding architecture.
        The \alpaka kernel provides about 20\% relative peak performance on all measured architectures.}
      \label{fig:performance_portability}
      \vspace{-1em}
\end{figure}

Figure~\ref{fig:performance_portability} shows the performance of the \alpaka tiling kernel executed on varying architectures relative to the theoretical peak performance of the corresponding architecture.
The kernel work division was selected in a way that provides good performance for the particular architecture.
CPU devices were accelerated by the \openmp~2 back-end, while NVIDIA devices were accelerated by the \cuda back-end.
The performance of all architectures lies around 20\% of the theoretical peak performance.
This shows that a single \alpaka kernel using all levels of the abstraction model together with optimized data access patterns is able to provide performance portability over various architectures.

\subsection{Real World Example}
 HASEonGPU is an open-source adaptive massively parallel multi-GPU Monte Carlo integration algorithm for computing the amplified spontaneous emission (ASE) flux in laser gain media pumped by pulsed lasers\footnote{\url{https://github.com/ComputationalRadiationPhysics/haseongpu}}.

The source code consists of about ten thousand lines of code and has been ported in three weeks by one person to \alpaka (HASEonAlpaka).
After the porting has been finished, HASEonAlpaka has successfully been executed on GPU and CPU clusters.

\begin{figure}[th]
  \centerline
      {\resizebox{0.5\textwidth}{!}{\includegraphics{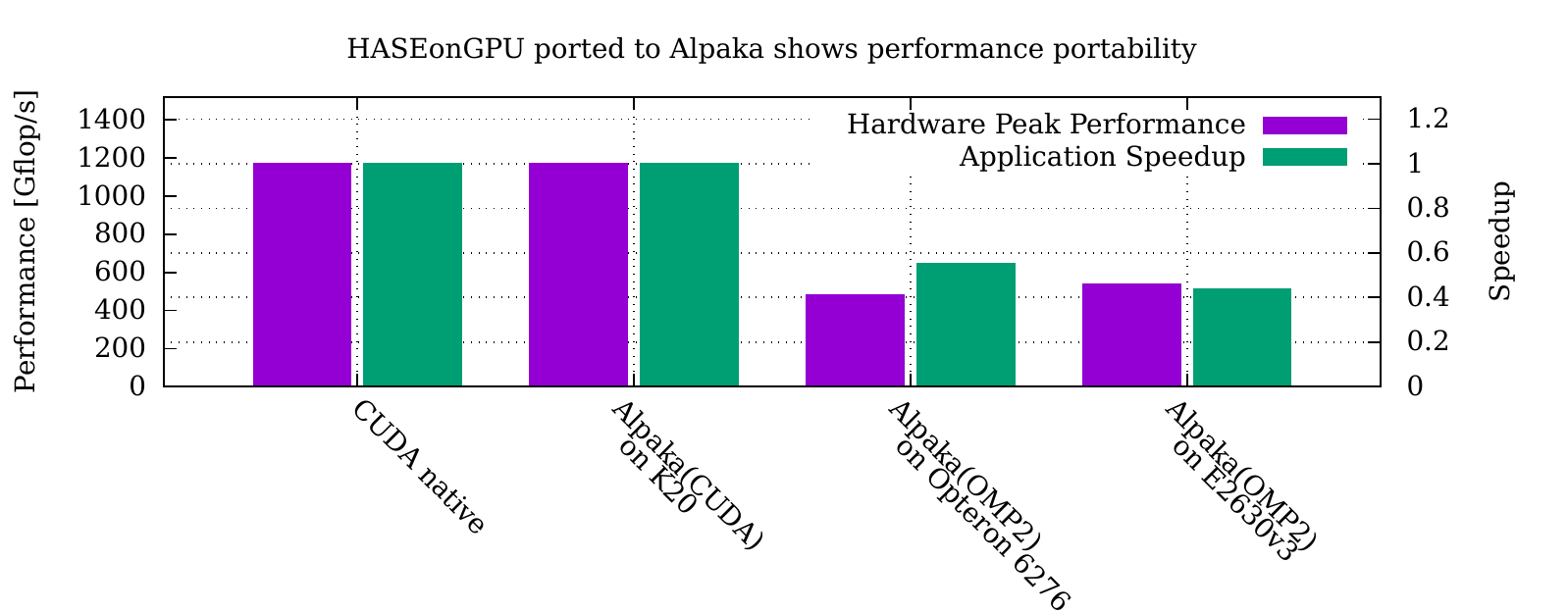}}}
  \caption{HASEonGPU was ported to \alpaka within three weeks by one person. The application shows almost perfect performance portability on all evaluated platforms.}
  \label{graphic:haseongpu}
        \vspace{-1em}
\end{figure}

Figure \ref{graphic:haseongpu} shows the relative speed of a HASEonAlpaka computation executed with identical parameters on different systems.
The original native CUDA version is used as the basis for comparison.
The \alpaka version using the \cuda back-end running on the same \nvidia K20 GK110 cluster as the native version does not show any overhead at all leading to identical execution times.

On the \intel Xeon E5-2630v3 and \amd Opteron 6276 clusters the \openmp~2  back-end without support for the not required thread level parallelism is used, i.e each block contains exactly one thread computing multiple elements.
This perfectly maps to the CPUs capabilities for independent vectorized parallelism and leads to very good results.
The nearly doubled time to solution on both, the \intel and \amd clusters, is on par with the halved double precision peak performance of those systems relative to the \nvidia cluster used as reference.

\section{Conclusion}
We have presented the abstract C++ interface \alpaka and its implementations for parallel kernel execution across multiple hierarchy levels on a single compute node.
We have demonstrated platform \emph{and} performance portability for all studied use cases.
A \emph{single source} \alpaka DGEMM implementation provides consistently 20\% of the theoretical peak performance on \amd, \intel and \nvidia hardware, being on par with the respective native implementations.
Moreover, performance measurements of a real world application translated to \alpaka unanimously demonstrated that \alpaka can be used to write performance portable code.

Performance portability, \emph{maintainability}, \emph{sustainability} and \emph{testability} were reached through the usage of C++ metaprogramming techniques abstracting the variations in the underlying architectures.
\alpaka code is \emph{sustainable}, \emph{optimizable} and easily extendable to support even more architectures through the use of C++ template specialization.
It is \emph{data structure agnostic} and provides a simple pointer based memory model that requires explicit deep copies between memory levels.

Future work will focus on including more \alpaka back-ends, e.g. for OpenACC and OpenMP 4.x target offloading and studying
performance portability for additional architectures (e.g \intel Xeon Phi and OpenPower) and applications.

\alpaka is an \emph{open-source} project and available in our \github repository\footnote{\url{https://github.com/ComputationalRadiationPhysics/alpaka}}.

\bibliography{citations}
\bibliographystyle{plain}


\end{document}